\documentstyle[12pt,psfig,aaspp4]{article}
\def\p0437{PSR J0437-4715}
\begin{document}
\centerline{\bf Title page}

\bigskip

\begin{tabular}{ll}

{\bf Title}:  & Time-scales of Radio Emission in \p0437 at 327 MHz \\
{\bf Short title}: & Time-scales in \p0437 \\
{\bf Name}: & M. Vivekanand \\
{\bf Address}:  & National Center for Radio Astrophysics  \\
		& Tata Institute of Fundamental Research  \\
		& Pune University Campus, P. O. Box 3 \\
		& Ganeshkhind, Pune 411007  \\
		& India. \\
{\bf Email}: & vivek@ncra.tifr.res.in \\
{\bf Status}: & Accepted for publication in Vol 543 (1 Nov 2000) of {\it The Astrophysical Journal} \\

\end{tabular}

\vfill
\eject
\title{\bf Time-scales of Radio Emission in \p0437 at 327 MHz}

\author{M. Vivekanand}

\affil{National Center for Radio Astrophysics, Tata Institute of Fundamental 
Research, Pune University Campus, P. O. Box 3, Ganeshkhind, Pune 411007, 
India.}

\begin{abstract}

%*****************************************************************************%
Time-scales of radio emission are studied in \p0437 at 327 MHz using almost 
half a million periods of high quality data from Ooty Radio Telescope. The 
radio emission in this milli second pulsar occurs on a short ($s$) time-scale 
of $\approx 0.026 \pm 0.001$ periods, and on a ($l$) time-scale that is much 
longer than the widths of the components of the integrated profile ($\approx 
0.05$ periods). The width of the $s$ emission increases with its increasing 
relative contribution to the total radio 
emission. This may provide constraints for the details of discharge of vacuum 
gaps above pulsar polar caps. The $s$ emission occasionally takes place in the 
form of intense spikes, which are confined to the main component of the integrated 
profile for 90\% of the time. The positions of spikes within a component of the 
integrated profile have no simple relation to the shape of that component. This 
may have impact on the interpretation of the integrated profile components in 
terms of independent regions of emission on the polar cap.

\end{abstract}


\section{Introduction}
%
%*****************************************************************************%
While it is becoming clear that the radio emission properties of normal and
milli second (ms) pulsars are broadly similar (Gil \& Krawczyk 1997; Jenet et 
al 1998; Kramer et al 1999), it is not yet clear exactly how similar they are. 
This is an important issue because of the three to four orders of magnitude 
difference in their periods and period derivatives. That a similar radio 
emission mechanism operates over such a wide range of the two most fundamental 
parameters of pulsars, is probably an important clue to the as yet unsolved 
problem of pulsar radio emission mechanism.

%*****************************************************************************%
This issue has been difficult to resolve since the radio emission from ms 
pulsars is very weak in comparison to that from normal pulsars, which are 
themselves considered weak radio emitters among cosmic sources. 
While several normal pulsars have been well studied for the details of their 
radio emission (see Manchester \& Taylor 1977), among ms pulsars \p0437 is one
of the few that are bright enough for such detailed studies (McConnell et al 
1996; Ables et al 1997 (paper I); Navarro et al 1997; Jenet et al 1998; 
Vivekanand et al 1998 (paper II)). This is the third paper in the series which 
discusses radio observations of \p0437 at 327 MHz using Ooty Radio Telescope 
(ORT), which has only a single polarization. This pulsar has a rotation period 
of $\approx$ 5.757 ms and is in a binary system of orbital period $\approx$ 
5.74 days. It is known to have radio emission in two modes -- one over times 
scales of $\approx$ 0.01 periods, and the other over $\approx$ 0.1 to 0.5 
periods. Paper I discussed one particular aspect of the former emission (spiky), 
while paper II discussed the overall properties of \p0437. Jenet et al 1998
study this pulsar in detail at $\approx$ 1400 MHz. This paper studies the time 
scales of radio emission in \p0437 at 327 MHz, using the auto-correlation 
function (ACF).

%*****************************************************************************%
Attempts have been made to model the integrated profile of \p0437 into 
independent components of emission (Gil \& Krawczyk 1997). In this paper the 
time-scales of radio emission are studied in the entire period, as well as in 
three major components of emission in the integrated profile; they are modeled 
as the Gaussian shown in Figure 1. The analysis was repeated for each of the 
three components by multiplying the flux density in each period by the Gaussian 
of the corresponding component. The ACF is independent of the amplitudes of the 
components.

\section{Auto-Correlation Function (ACF)}

%*****************************************************************************%
The details of data acquisition and pre-analysis are described in papers I and
II. Briefly, ORT was used to obtain 436,500 periods of high quality data on 
\p0437, sampled at 0.1024 ms. The data were collected in 36 files each 
containing 12,125 useful periods of continuous data (paper II analyzed 34 of
these files). Since the pulsar's rotation period is continuously Doppler 
shifted, the data was re-sampled at $\approx$ 0.1028 ms, depending upon the 
average period of the pulsar in each file, so that there are exactly 56 time 
samples in each period. The data in each file were shifted so that all 36 
integrated profiles were aligned, correct to half a sample. Thus the radio 
emission properties can be studied as a function of ``phase'' or ``longitude'' 
within each period, across all files (Vivekanand et al 1998).

%*****************************************************************************%
Figures 2 and 3 describe the method of obtaining the time scales of radio 
emission in \p0437. First, the data of each period were centered in an array of 
length 64 samples (top frame of Figure 2). Then they were auto-correlated after 
weighting by a Hamming function. The resulting ACF is shown in the bottom frame 
of Figure 2, as a function of the time delay $t$. Since the ACF is an even 
function of $t$, it was fit to a curve of the form

\begin{equation}
\mathsf{ACF} = k_0 + k_1 t^2 + \sum_{i = 1}^{N_g} a_i \exp - \left [ 
	\frac{t}{b_i} \right ]^2.
\end{equation}

\noindent
%*****************************************************************************%
The analysis used routines from Numerical Recipes (Press et al 1989), after 
modifying and debugging them for double precision arithmetic. The ACF for each 
of the 436,500 periods was fit to four variations of equation 1: $N_g = 2$ 
and $N_g = 3$, each of which without and with the quadratic term $k_1$. This 
was found sufficient to model more than 75\% of the data.

%*****************************************************************************%
For each of the four variations, the parameters of the fit were constrained as 
follows (provided that the basic non-linear fit converged): (1) The root mean 
square (rms) error $\sigma_f$ on any parameter $f$ in equation 1 should be 
less than 1.0 fractionally, i.e., $\sigma_f / f < 1.0$; (2) The amplitudes $a$ 
of the Gaussian should be positive, otherwise the derived time-scale will not
make sense, and should be less than 1.0 + $\sigma_a$ (since the peak of the ACF 
is normalized to 1.0 by definition); (3) For the same reason, the estimated 
value of the ACF at $t$ = 0.0, which is $k_0 + \sum_{i = 1}^{N_g} a_i$, must be 
equal to 1.0, consistent with its rms error; (4) The absolute value of the 
coefficient $k_0$ must be less than 0.1; and (5) The term $\left \vert k_1 t^2 
\right \vert $ must not contribute more than 0.1 (in magnitude) to the ACF 
at the extreme values of the abscissa ($t = \pm 0.5174$ periods) in the bottom
frame of Figure 2. 

%*****************************************************************************%
For each value of $N_g$, the results of the corresponding two variations 
(without and with $k_1$) were combined in the following manner: (1) When 
neither variation was able to fit the ACF, that period was set aside for later 
fitting, or given up altogether; (2) When only one variation fit the ACF, that 
solution was adopted; (3) When both variations fit the ACF, the following was 
done: 

\begin{itemize}

%*****************************************************************************%
\item The Gaussian were sorted in the increasing order of width ($b$) for 
proper comparison between the two variations.

\item All parameters except $k_1$ in equation 1 were checked for consistency 
between the two variations, by the formula 

\begin{equation}
 \left \vert \frac{f_1 - f_2}{\sqrt{\sigma_{f_1}^2 + \sigma_{f_2}^2}} \right 
 \vert < 3.0.
\end{equation}

\noindent
%*****************************************************************************%
where the subscripts $1$ and $2$ refer to the above two variations, respectively.

%*****************************************************************************%
\item If all above parameters were consistent in the two variations, then the 
results with $k_1$ fitted were chosen if $k_1$ was significant ($\sigma_{k_1} 
/ k_1 < 1.0$); otherwise the results without $k_1$ fitted were chosen. 

%*****************************************************************************%
\item If even one of the above parameters was inconsistent in the two 
variations, then the results with the lower $\chi^2$ were chosen, by using 
equation 2 with $f$ replaced by $\chi^2$, and with 3.0 replaced by 2.0. A 
minimum $\chi^2$ criterion should be applied with caution in this problem, 
because of fits such as that in Figure 3, which are considered good for 
obtaining the time-scale of emission, although the $\chi^2$ is high.

\end{itemize}

%*****************************************************************************%
The results with $N_g = 3$ were utilized for only those periods that could not 
be fit with $N_g = 2$. Finally, only those results were retained whose $\chi^2$ 
deviated from the mean value in each file by less than three standard 
deviations.

\subsection{Discussion of the Model Fits}

%*****************************************************************************%
The best fit parameters for the ACF in Figure 2 are: $N_g = 2$, $a_1 = 0.2541 
\pm 0.0031$, $b_1 = 0.0297 \pm 0.0005$, $a_2 = 0.8190 \pm 0.0037$, $b_2 = 
0.2049 \pm 0.0010$, $k_0 = -0.0728 \pm 0.0041$, $k_1 = 0.2509 \pm 0.0314$, the 
rms deviation of the ACF about the fitted curve being $0.0138$. The rms errors 
were obtained in the usual manner using the covariance matrix, the rms 
deviation of the ACF, and the number of degrees of freedom. The best fit 
parameters in Figure 3 are: $N_g = 2$, $a_1 = 0.5117 \pm 0.0109$, $b_1 = 0.0159 
\pm 0.0005$, $a_2 = 0.5068 \pm 0.0054$, $b_2 = 0.2366 \pm 0.0035$, $k_0 = 
-0.0171 \pm 0.0037$, $k_1$ not being fit, the rms deviation of the ACF being 
$0.0504$. The latter rms deviation is significantly higher due to the double 
pulsed structure in the top frame of Figure 3, which creates the symmetric 
peaks in the ACF at around $t = \pm 0.1$ periods in the bottom frame, which 
are fit only in a mean sense. This will slightly over estimate the width $b_2$ 
for the wider component of the ACF.

%*****************************************************************************%
The above analysis fitted 81.54\% of the periods without any component 
weighting, and 76.69\%, 84.66\% and 76.00\% of the periods in the three 
components of the integrated profile in Figure 1, respectively. This was 
considered sufficient because of the extremely large number of periods 
available (436,500) to begin with.  The majority of the periods required only 
two time-scales of emission for proper representation; the fraction of the 
fitted periods that required $N_g = 3$ was 0.24\%, 10.91\%, 1.25\% and 1.67\% 
in the above four cases, respectively.

%*****************************************************************************%
It is not possible to visually compare the curve fits with the computed ACF for
all periods. This was done for 4800 randomly chosen periods, or $\approx$ 1.1\% 
of the total periods. In addition, 20 periods having the lowest $\chi^2$ and 20 
periods having the highest $\chi^2$ were also examined visually in four of the
36 data files (that were also chosen randomly). Only one or two fits 
were found to be unsatisfactory. Assuming that this number is 3, the maximum 
fraction of misfit periods one expects is $3 / 4800 \approx 0.0063$\%. However
fluctuations in this fraction will lead to a maximum fraction of misfit periods
of $\left ( 3 + 5 \times \sqrt{3} \right ) / 4800 \approx 0.24$\%, at the five
standard deviation confidence level; this takes into account the fact that only 
$\approx 1.1$\% of the total number of periods were tested. This is unlikely to 
influence the results of this paper. This random checking was repeated for each 
of the three components of the integrated profile in Figure 1, with similar 
results.

\section{Duration of Time-Scales}

%*****************************************************************************%
Figure 4 shows the normalized probability density $p(a, b)$ of occurrence of 
$a$ and $b$ for the entire integrated profile and for the three components of 
the integrated profile in Figure 1.  The estimated values of $a$ (or $b$) were 
assumed to be Gaussian random variables with standard deviation $\sigma_a$ (or 
$\sigma_b$), which are the rms errors obtained by curve fitting the ACF. 
Assuming that the errors $\sigma_a$ and $\sigma_b$ are uncorrelated, $p(a, b)$ 
can be defined as

\begin{eqnarray}
\nonumber &p(a, b) &= \frac{1}{2 \pi N_P} \sum_{i = 1}^{N_P} \\
\nonumber & & \left [ \frac{1}{\sigma_{a_s} \sigma_{b_s}} \exp - \left \{ \left ( \frac{a - 
a_s}{\sigma_{a_s}} \right )^2 + \left ( \frac{b - b_s}{\sigma_{b_s}} \right )^2 
\right \} \right ]_i \\
&  &+ \left [ \frac{1}{\sigma_{a_l} \sigma_{b_l}} \exp - \left \{ \left ( \frac{a - 
a_l}{\sigma_{a_l}} \right )^2 + \left ( \frac{b - b_l}{\sigma_{b_l}} \right )^2 
\right \} \right ]_i
\end{eqnarray}

\noindent
%*****************************************************************************%
where the subscripts $s$ and $l$ refer to small and large time-scales, 
respectively, and the sum is taken over all valid periods $N_P$. The 
distribution of the third time-scale is similar to that of $l$, and they are 
insignificant in number anyway; so they have not been plotted in Figure 4. Note
that any correlation observed in the function $p(a, b)$ refers to that between
the mean values of $a_s$ and $b_s$, or $a_l$ and $b_l$.

%*****************************************************************************%
Table 1 shows the mean values of the widths $b$ of the above three time-scales,
estimated in the four plots of Figure 4. The rms errors for the third 
time-scale are significantly higher than the rest because of the much smaller
numbers. It is easy to verify that the longer time-scales ($l$ and $3$) scale 
roughly as the width of the corresponding component of the integrated profile, 
while the shorter time-scale ($s$) is almost constant in the three components.

%*****************************************************************************%
The amplitudes $a$ of the $s$ and $l$ time-scales are anti-correlated with each
other; this is expected since the peak ACF is normalized to 1.0, and the third 
time-scale contributes insignificantly to the ACF. The amplitudes $a$ are a 
rough measure of the relative energy in those time-scales. It is difficult to 
obtain from them the exact energy in each time-scale of emission in \p0437, due
to the behavior of the ACF as in Figure 3.

%*****************************************************************************%
Further, the amplitudes $a$ and widths $b$ are correlated for the short 
time-scale ($s$) emission, while they are not correlated for the long time-scale 
($l$) emission. This is evident in all four plots of Figure 4. To study this 
quantitatively, horizontal cuts were taken in plots (a) and (c) of Figure 4, the 
former for the mean effect, and the latter for the maximum effect. The $p(a, b)$ 
cut at each $a$ was modeled as two Gaussian, to represent the two time-scales. 
The positions $<b>$ of these Gaussian were fit to a straight line as a function 
of $a$ (these are shown in Figures 4a and 4c):

\begin{equation}
<b> = \alpha + \beta a.
\end{equation}

\noindent
%*****************************************************************************%
In Figure 4a, $\alpha = 0.0175 \pm 0.0002$ and $\beta = 0.0135 \pm 0.0005$ for 
the $s$ emission, while $\alpha = 0.2129 \pm 0.0027$ and $\beta = 0.0060 \pm 
0.0040$ for the $l$ emission. The corresponding numbers in Figure 4c are 
$\alpha = 0.0189 \pm 0.0004$ and $\beta = 0.0159 \pm 0.0008$ for the $s$ 
emission, and $\alpha = 0.0621 \pm 0.0005$ and $\beta = -0.0023 \pm 0.0009$ 
for the $l$ emission. In both plots $\beta$ is much larger for the $s$ 
emission than for the $l$ emission, for which it is almost negligible; it is 
also much more significant relative to its rms error. This verifies the above 
claim.

%*****************************************************************************%
An alternate fit was also tried by incorporating a quadratic term $\gamma a^2$ 
into equation 4. For the $s$ emission the earlier trend was verified, while for 
the $l$ emission it gave a curvature in the opposite sense, which is also 
obvious by visual inspection of Figure 4. It may be tempting to claim here that 
the $a$ and $b$ for the $l$ emission are actually anti-correlated. However, 
this could also be due to the fact that our curve fitting to the ACF gave 
biased (higher) values of $b$ in data such as that shown in Figure 3; and this 
effect appears to be larger for lower $a$ (i.e., for weaker $l$ emission). 
Therefore the possible anti-correlation of $a$ and $b$ in the $l$ emission has 
to be justified with better data or with more sophisticated analysis. Note that
such a problem does not occur for the $s$ emission.

%*****************************************************************************%
The trend seen in Figure 4 was noticed in all 36 individual files of data. It
was verified visually for some of the strongest ``spikes'' of radio emission. It
is also consistent with the claim of Jenet et al 1998 that they ``observe a 
significant inverse correlation between pulse peak and width'' for the spiky 
emission (see below). Therefore this appears to be a genuine feature of radio 
emission in \p0437. This result is also independent of the type A and type B 
integrated profiles discussed in Figure 8 of Vivekanand et al 1998.

\section{Distribution of Spikes}

%*****************************************************************************%
Ables et al 1997 studied the distribution of the spiky emission from \p0437 in
the ``phase'' or ``longitude'' space. They claim that ``The spikes are 
observed almost exclusively in a 10$^\circ$ phase window centered on the main 
pulse.  Within that window the phase distribution has a periodic variation''. 
The latter claim has been found invalid by Jenet et al 1998. Ables et al 1997 
claimed a very high resolution in the phase space; this section attempts a more 
modest (and probably what should have been the more preliminary) study of the 
problem.

%*****************************************************************************%
The method of analysis is similar to that described earlier, except that (1) 
only those periods were chosen which had flux density peaks higher than a 
threshold value of 10.0 (in arbitrary units) in the data file with the highest 
average energy per time sample $< E >_{m} = 0.6622$; for any other file the 
threshold was determined by $10.0\  \times < E > /\ < E >_{m}$; (2) the curve 
fitting was done to the flux density of the highest peak and not to the ACF 
(i.e., to the top frame in Figures 2 and 3); and (3) the position of the peak 
was also fit. This was done for components 1 and 2 of the integrated profile, 
with the following threshold values for the most luminous data file: 7.5, 10.0, 
12.5 and 15.0 (at this level of threshold, no spikes were found in component 3). 
About 90\% of the spikes occurred in component 2, in which the number of 
periods with spikes above the four thresholds were 10,853, 2735, 800 and 222, 
respectively. This implies that the probability of occurrence of spikes 
decreases $\approx$ exponentially with their peak flux density. This result 
also holds for component 1.

%*****************************************************************************%
Figure 5 shows the probability  distribution of the peak flux densities and
positions of the spikes in component 2; the technique is similar to that used 
in obtaining Figure 4, where one assumed the abscissa and the ordinate to be
distributed as Gaussian random variables. Integrating this plot vertically one 
obtains the probability distribution of the positions of the spikes within
component 2.  Comparing this with the component 2 profile in Figure 1, it is 
clear that the probability of occurrence of the spikes in pulsar rotation 
``phase'' is not simply related to the shape of the integrated profile in that 
range of phase. It is also not simply related to the integrated profile formed
by the spikes alone, which is only marginally different from Figure 1. By 
visual inspection it appears that this distribution can be modeled in several 
ways, including possibly as a Gaussian like peak on top of a plateau that is 
tapering down towards increasing abscissa. It is probably wise to await more 
or better data before doing rigorous curve fitting to this probability 
distribution.

%*****************************************************************************%
Figure 6 shows the same information for spikes occurring in component 1 of the
integrated profile; the probability distribution looks similar to that in 
Figure 5, except that it is much more noisy, since it contains 303 periods 
only. These results are independent of the type A and type B integrated 
profiles discussed in Figure 8 of Vivekanand et al 1998.

%*****************************************************************************%
This section can not comment on the veracity of the Ables et al 1998 result
that has been contested, since the technique used here can not achieve the 
phase resolution claimed by them. The phase resolution here is determined by
the rms errors obtained by the curve fitting procedure.

%*****************************************************************************%
Figure 7 shows the probability of occurrence of the peak fluxes and widths of 
the spikes in component 2 of the integrated profile; it verifies the inverse
correlation seen in Figure 8 of Jenet et al 1998 (which, the reader is reminded,
was observed at $\approx$ 1400 MHz). However, Figure 7 also shows that the 
average energy in a spike is an increasing function of the width, consistent 
with the results of Figure 4. This implies that the inverse correlation between 
peak fluxes and widths is such that their product shows a positive correlation.

\section{Discussion}

%*****************************************************************************%
First, one will compare the results on \p0437 obtained by Jenet et al 1998 at
1380 MHz, and by Vivekanand et al 1998 (paper II) and this work at 327 MHz. The 
former use coherent dedispersion to obtain a typical time resolution of 2.56 
$\mu$ sec, with dispersion smearing of 3.26 $\mu$ sec, while the latter have a 
time resolution of 102.8 $\mu$ sec, which is several times worse. The former 
conclude that \p0437 shows neither systematic drifting sub-pulses nor the 
nulling phenomenon at 1380 MHz, and note the broad spectral feature in the 
fluctuation spectrum. The latter confirm these results at 327 MHz, and study 
them in much greater detail. The former do not test for long time scale (of the 
order of several tens or hundreds of periods or more) flux variations, while 
the latter show that \p0437 has such flux variations at 327 MHz, although this 
must be verified with full polarization data. The pulse energy distributions 
are similar at both radio frequencies, with the conclusion that there are no 
giant pulses in \p0437.  Both works conclude that there is no emission in \p0437 
at very short time scales, which is canonically known as ``micro-structure'' in 
pulsars; in this regard the current work is limited by the relatively large 
sampling interval. The former show that the quasi periodic phase modulation of 
the spiky emission in \p0437, that was claimed by Ables et al 1997 (paper I), 
does not exist at 1380 MHz; the latter do not have the time resolution to verify 
this. The former show that the peak fluxes of the spikes are anti-correlated 
with their widths; the latter confirm this. Based on this the former claim that 
the pulse energies are more or less constant, while the latter show that they 
are positively correlated with the pulse widths.

%*****************************************************************************%
The spiky emission of \p0437 must be studied keeping in mind the giant pulse 
phenomenon observed in the ms pulsar PSR B1937 +21 (Sallmen \& Backer 1995), 
and the normal Crab pulsar. Until all radio pulsars within a period range are 
studied systematically for spiky emission, which has probably not occurred so 
far, it is not possible to claim whether the rotation period is the main 
parameter that determines the spiky emission from radio pulsars. Such an 
exercise will also throw light on whether the low period (0.253 s) but normal 
pulsar PSR B0950+08, that is known to have impulsive emission (Manchester \& 
Taylor 1977), falls roughly in the same category as \p0437.

%*****************************************************************************%
The time-scale of the $l$ emission is expected to roughly scale as the width 
of the component of the integrated profile that it occurs in, due to the 
multiplication by the corresponding Gaussian before the ACF is computed, if it
is a genuinely long time-scale emission, much longer than the widths of the
components. Conversely, the width of the $s$ emission is expected to be 
independent of the component widths if it is much smaller than them. The actual
time-scale of the $s$ emission is unlikely to be be shorter than that mentioned 
in Table 1 because of the sampling interval (0.1024 ms) of this data (Jenet et 
al 1998).

%*****************************************************************************%
The result that the $s$ time-scale is an increasing function of its relative 
contribution to the integrated flux density, should be contrasted with the 
behavior reported in normal pulsars (Manchester \& Taylor 1977), as well as in 
\p0437 itself, where the peaks of the sub-pulses are anti-correlated with their 
widths. At first glance the latter result might imply that the integrated 
fluxes of sub-pulses are roughly independent of their widths, which is not true
at least in \p0437. So it is important not to conclude one from the other. To 
the best of this author's knowledge Figure 4 has not been estimated for 
sub-pulses of normal pulsars. Two points need to be clarified here. First, the 
spikes reported here have widths (in ``phase'' space) comparable to those of 
sub-pulses in normal pulsars, and not those of micro-structure (also see Jenet 
et al 1998). Second, the above comparison is justified only if the ``relative 
contribution to the integrated flux'' also implies the ``integrated flux'' in 
the average sense, which is expected from a statistical viewpoint unless 
correlations are observed to the contrary. Further, the visual inspection 
mentioned earlier, as well as Figure 7, confirms such conjecture.

%*****************************************************************************%
It will be interesting to verify Figure 4 of this paper in the data of Jenet 
et al 1998. If verified, the positive correlation between the energies and 
widths of the spikes would be a broad band phenomenon, which behaves more like 
micro-structure, rather than sub-pulses in normal pulsars!

%*****************************************************************************%
The above might have important implications for the details of the discharge 
of the vacuum gap above pulsar polar caps (Ruderman \& Sutherland 1975). Let us 
assume that the basic pulsar radio emission occurs in terms of sparks, and 
several contiguous sparks make a sub-pulse. If the energy in a sub-pulse is 
independent of its width, it could imply that the discharge process is 
operating at a maximum threshold level; in other words, once a series of 
sparks is initiated they probably extract a fixed (maximum) amount of energy 
from the vacuum gap. On the other hand, if the energy in a sub-pulse increases
with increasing width, it might imply that the discharge process is operating
at a level much lower than the maximum threshold; in other words, there is
much more energy available than is usually extracted by the sparks, and more
contiguous sparks imply more energy. Finally, if the energy in a sub-pulse
decreases with increasing width, it might imply a high level of electric
conductivity on the polar cap (transverse), which is not expected in normal
pulsars due to the very huge magnetic fields, but may be plausible in ms 
pulsars. A further implication might be that the transverse development of the 
series of sparks on the polar cap (which is what the width essentially refers 
to) may or may not impede the vertical development of the sparks above the 
polar cap (which is what the energy essentially refers to). This may also be 
related to the competition of sub-pulse energies noticed in the drifting
pulsar PSR B0031 -07 by Vivekanand \& Joshi 1999. Such might be the kind of 
implications which might provide tight constraints on the details of the 
discharge of the vacuum gap.

%*****************************************************************************%
That the spiky emission is almost absent from the third component of the 
integrated profile of \p0437 further supports the special behavior of this 
component reported in Vivekanand et al 1998.

%*****************************************************************************%
In principle, the shape of the integrated profiles of radio pulsars should be 
determined by the convolution of the distribution of sub-pulse positions with
the distribution of sub-pulse energies, taking into account the shapes of
sub-pulses, the random variations in them, etc. While mathematically there may 
be no further constraints on these two distributions, physically one would expect 
them to be simply related to the relevant component of the integrated profile.
This is certainly not the case for the spiky emission in \p0437. This raises
questions such as, what exactly are the components of integrated profiles?
Are they really individual beams of emission (Gil \& Krawczyk 1997)? Is this 
picture justified in \p0437 where the $l$ emission is much wider than the 
components of the integrated profile? Clearly polarization data is an important 
independent input for this problem.

%*****************************************************************************%
Another independent method of resolving the integrated profile into {\it 
independent regions of emission on the polar cap} is proposed here. It assumes 
that the pulsar flux varies independently in each of these regions of emission.  
Then the cross-correlation of pulsar data at all pairs of phases (longitudes) 
within the integrated profile can be modeled, and optimized in terms of the 
minimum number of independent regions required for the model. To implement this, 
the pulsar data must show significant cross-correlation, which \p0437 does. 
This is a difficult work and is currently in progress.

%*****************************************************************************%
The main results of this paper are:

\begin{enumerate}

%*****************************************************************************%
\item \p0437 emits radio waves at 327 MHz on two time-scales: the ($s$) time-scale 
of $\approx 0.026 \pm 0.001$ periods, that is much smaller than the widths of 
the components of the integrated profile, and the ($l$) time-scale that is much 
longer than the component widths (Table 1). The $s$ emission occasionally takes 
place in the form of intense spikes.

%*****************************************************************************%
\item The time-scale of the $s$ emission is positively correlated with its 
relative contribution to the total radio emission in a given period. It is 
necessary to find out if the same occurs in normal pulsars also. This might
constrain the details of discharge of vacuum gaps above pulsar polar caps.

%*****************************************************************************%
\item The spiky emission in \p0437 is confined to the main component of the 
integrated profile (component 2 in Figure 1) for 90\% of the time, and to 
component 1 for the rest 10\% of the time; it rarely occurs in component 3, 
which supports earlier claims of this component's special behavior (Vivekanand
et al 1998).

%*****************************************************************************%
\item The distribution of the positions of the spikes within a component of the
integrated profile has no simple relation to the shape of that component. This 
raises questions concerning the meaning of the integrated profiles components 
in terms of independent regions of emission on the polar cap.

\end{enumerate}

\vfill
\eject

\noindent
{\bf References}

\begin{enumerate}

%*****************************************************************************%
\item Ables, J. G., McConnell, D., Deshpande, A. A. \& Vivekanand, M. 1997, ApJ, 
475, L33

\item Gil, J. \& Krawczyk, A. 1997, MNRAS, 285, 561

\item Jenet, F. A., Anderson, S. B., Kaspi, V. M., Prince, T. A. \& Unwin, S. C.
1998, ApJ, 498, 365

\item Kramer, M., Lange, C., Lorimer, D. R., Backer, D. C., Xilouris, K. M.,
Jessner, A. \& Wielebinski, R. 1999, ApJ, 526, 957

\item Manchester, R. N. \& Taylor, J. H. 1977, Pulsars, W. H. Freeman \& Co.

\item McConnell, D., Ables, J. G., Bailes, M. \& Erickson, W. C. 1996, MNRAS,
280 331

\item Navarro, J., Manchester, R. N., Sandhu, J. S., Kulkarni, S. R. \& Bailes, M.
1997, ApJ, 486, 1019

\item Press, W. H. Flannery, B. P., Teukolsky, S. A. \& Vetterling, W. T.  1989, 
Numerical Recipes (Cambridge: Cambridge Univ. Press)

\item Ruderman, M. A. \& Sutherland, P. G. 1975, ApJ, 196, 51

\item Sallmen, S. \& Backer , D. C. 1995, Astron. Soc. Pac. Conference Series,
72, 340

\item Vivekanand, M., Ables, J. G. \& McConnell, D. 1998, ApJ, 501, 823

\item Vivekanand, M. \& Joshi, B. C. 1999, Apj, 515, 398

\end{enumerate}

\bigskip

\begin{deluxetable}{ccccc}
\large
%*****************************************************************************%
\tablecaption{The first three columns contain the mean value of the width $b$, 
defined as $\sum_{i = 1}^{N_P} b_i / N_P$, and its rms error defined as $\sqrt{ 
\sum_{i = 1}^{N_P} \left [ \sigma_{b_i} / N_P \right ]^2}$, for the three 
time-scales of radio emission. The four rows correspond to the four plots of 
Figure 4. For comparison, the widths of the three components of the integrated 
profile shown in Figure 1 (modeled as Gaussians) have been reproduced in the 
last column.}
\tablehead{\colhead{} & \colhead{$s$}  & \colhead{$l$}  &  \colhead{$3$}  & \colhead{}
}
\startdata
(a) & $0.0234 \pm 0.0023$ & $0.2020 \pm 0.0061$ & $0.2341 \pm 0.0475$  &  \nl
(b) & $0.0255 \pm 0.0016$ & $0.0995 \pm 0.0030$ & $0.1080 \pm 0.0018$  & 0.078 \nl
(c) & $0.0271 \pm 0.0005$ & $0.0611 \pm 0.0011$ & $0.0684 \pm 0.0066$  & 0.051 \nl
(d) & $0.0263 \pm 0.0007$ & $0.0561 \pm 0.0013$ & $0.0596 \pm 0.0064$  & 0.044 \nl
\enddata
\end{deluxetable}

\vfill
\eject

\clearpage
\begin{figure*}
\unitlength=1.0cm
\centering
\begin{picture}(17,17)(0,0)
\psfig{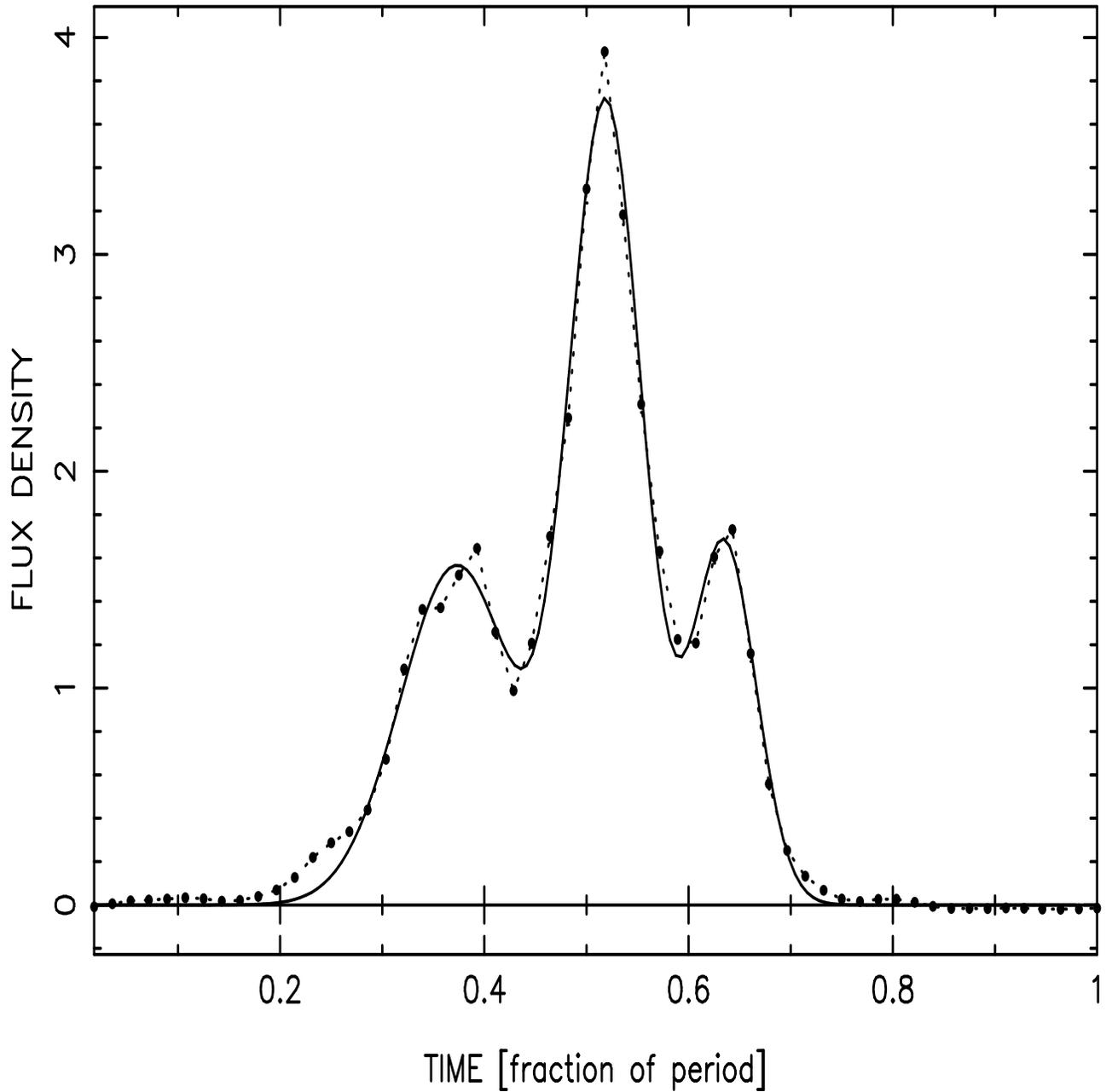}
\end{picture}
\caption{Integrated Profile of \p0437 at 327 MHz using 436,500 periods (dots). The
abscissa is time in units of the rotation period of \p0437, while the ordinate
is the flux density in arbitrary units. It has been resolved into three major
components of emission, each represented by a Gaussian, whose sum is the smooth
curve. The Gaussians are centered at 0.373, 0.519 and 0.635 periods, having widths
0.078, 0.051 and 0.044 periods, respectively. The abscissa is also known as the
``phase'' or ``longitude'' within the period.}
\end{figure*}

\vfill
\eject
\clearpage
\begin{figure*}
\unitlength=1.0cm
\centering
\begin{picture}(17,17)(0,0)
\psfig{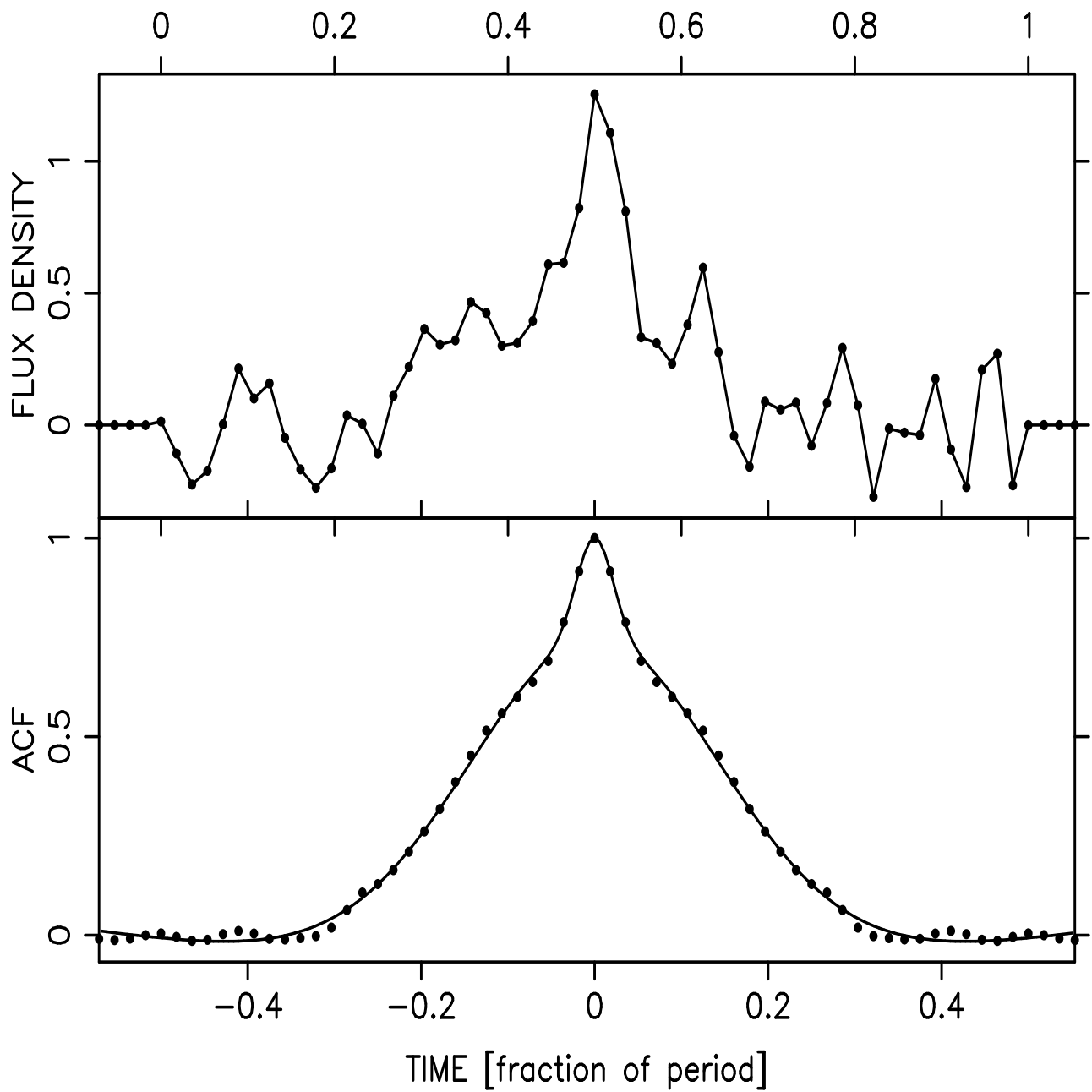}
\end{picture}
\caption{{\bf Top frame}: Flux density (in arbitrary units) as a function of time
(in fraction of period) for the 12$^{\mathsf{th}}$ period in the data file 
observed on 1995 Mar 7 starting at 16:04:17 UT (observation code 50661604). The 
duration of the synthesized samples is 0.10281556 ms. The 56 samples of data are 
centered in an array of length 64 samples (for doing FFT). {\bf Bottom frame}: The 
ACF of the data in the top frame, as a function of the time delay $t$. The dots 
represent the computed ACF, while the smooth curve is the best fit discussed in 
the text. The number of degrees of freedom is 32 (not 64) minus the number of 
parameters fit.}
\end{figure*}

\vfill
\eject
\clearpage
\begin{figure*}
\unitlength=1.0cm
\centering
\begin{picture}(17,17)(0,0)
\psfig{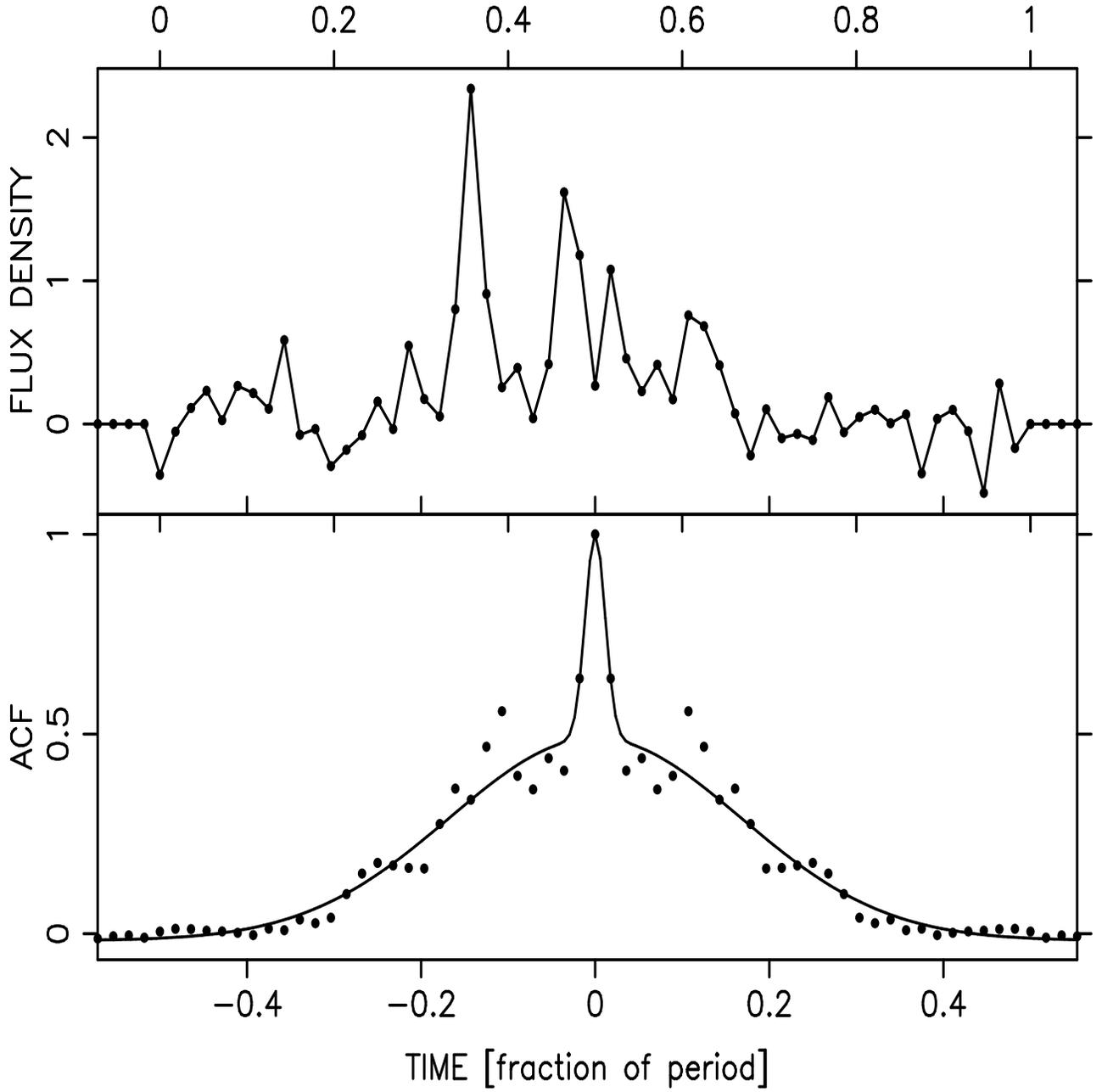}
\end{picture}
\caption{Same as in Figure 2, except that it is the 27$^{\mathsf{th}}$ period of 
the same data file.}

\end{figure*}

\vfill
\eject
\clearpage
\begin{figure*}
\unitlength=1.0cm
\centering
\begin{picture}(17,17)(0,0)
\psfig{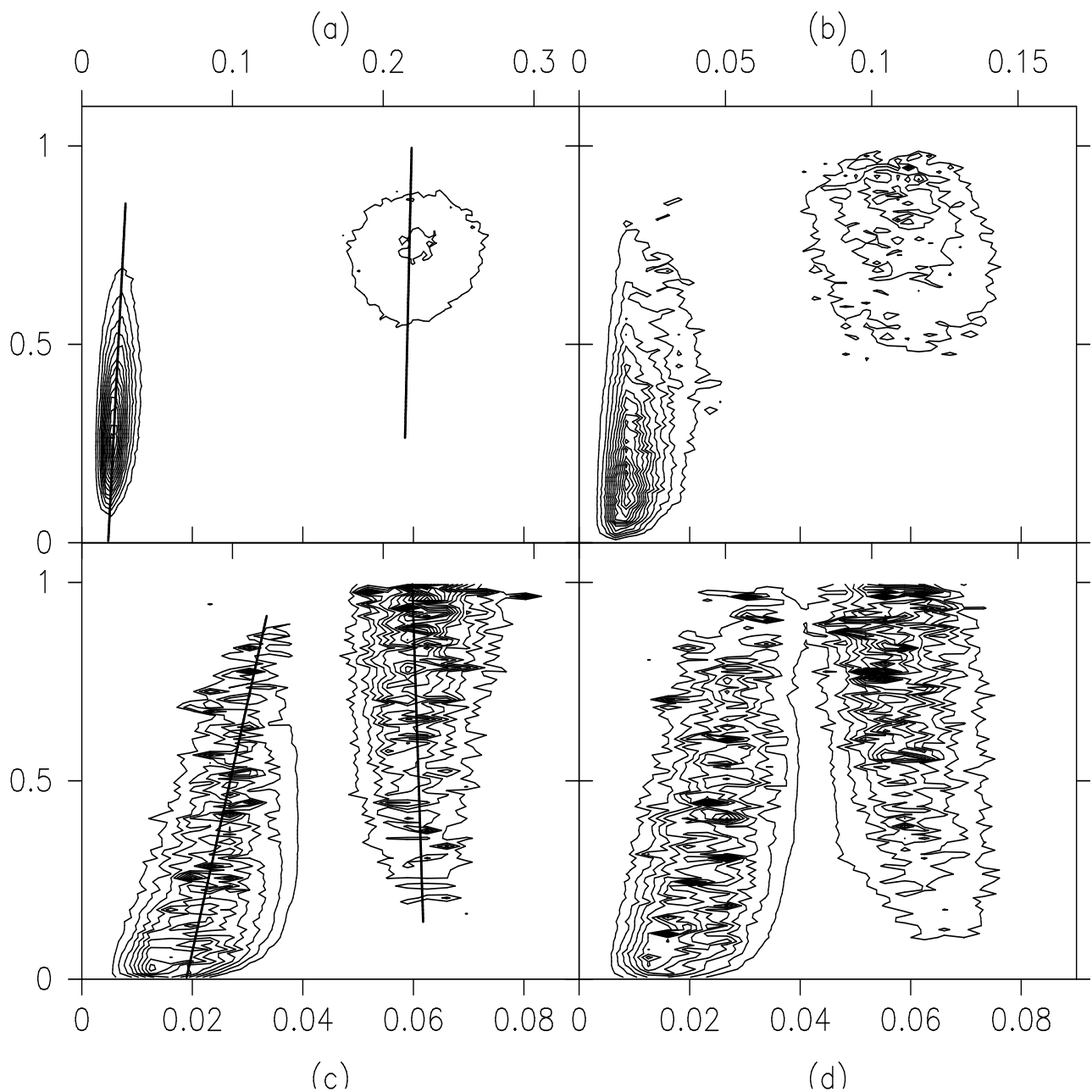}
\end{picture}
\caption{Contour plots of $p(a, b)$ as a function of amplitudes $a$ (ordinate) 
and widths $b$ (abscissa) defined in equation 1. Plot (a) belongs to the entire 
integrated profile, while plots (b), (c) and (d) belong to the three components 
of the integrated profile defined in Figure 1, respectively. The total abscissa 
range (0.0 to 1.0 periods) and total ordinate range (0.0 to 1.0) were divided 
into 100 bins. The contours are 15 in number equally spaced in the $p(a, b)$ 
space. The thick straight lines in plots (a) and (c) show the trend of the 
corresponding probability density in the $a$ -- $b$ plot.}
\end{figure*}

\vfill
\eject
\clearpage
\begin{figure*}
\unitlength=1.0cm
\centering
\begin{picture}(17,17)(0,0)
\psfig{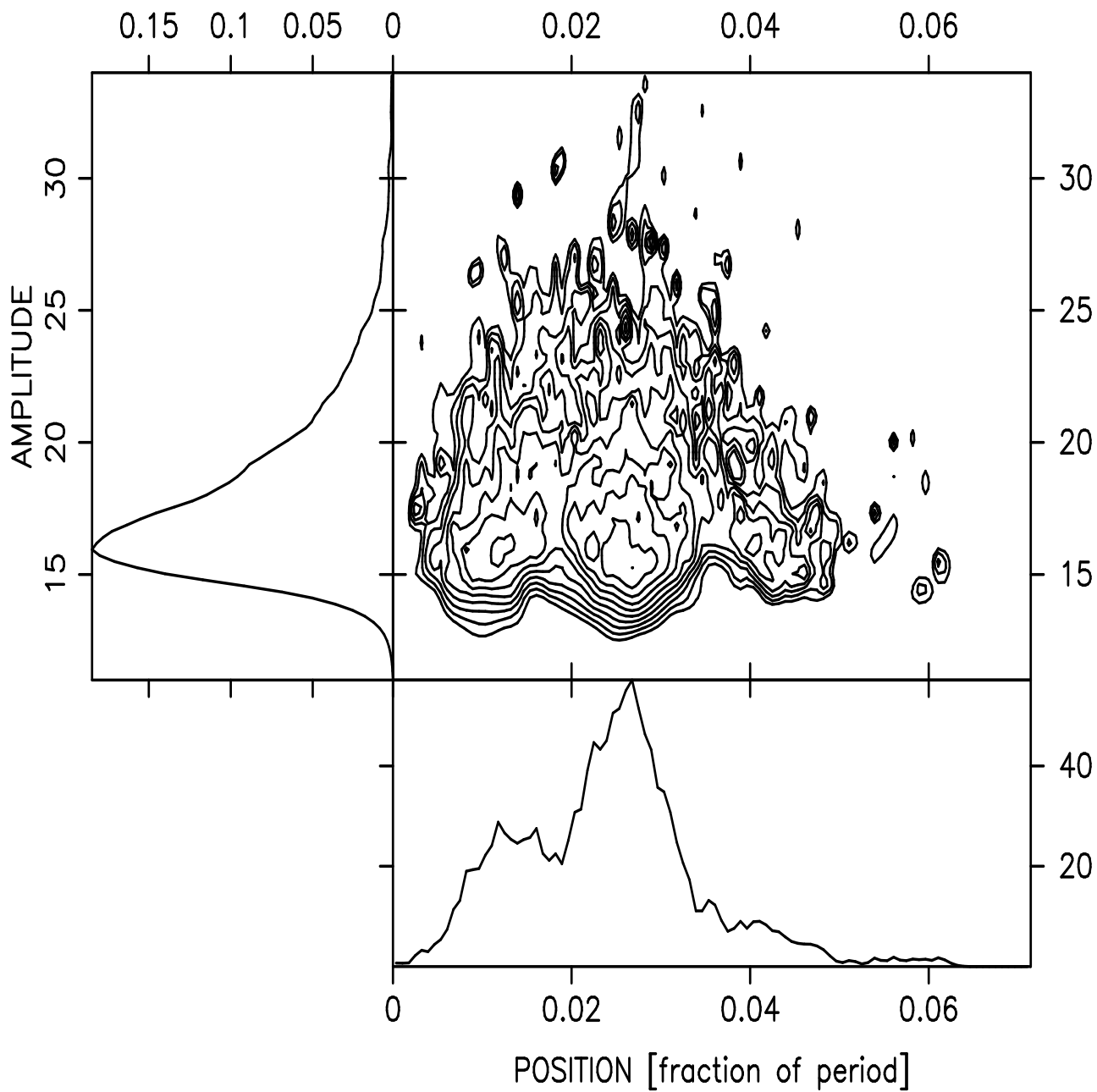}
\end{picture}
\caption{Probability density function of the peak flux density (ordinate) of the 
spikes versus their position (abscissa) in component 2 of the integrated profile
shown in Figure 1. For reference the peak of component 2 lies at 0.027 on the 
abscissa. Fifteen contours are equally spaced in the logarithm of the 
probability. One dimensional probability distributions when the above plot is 
integrated horizontally and vertically are also shown. The threshold flux 
density for the most luminous data file was 10.0.}
\end{figure*}

\vfill
\eject
\clearpage
\begin{figure*}
\unitlength=1.0cm
\centering
\begin{picture}(17,17)(0,0)
\psfig{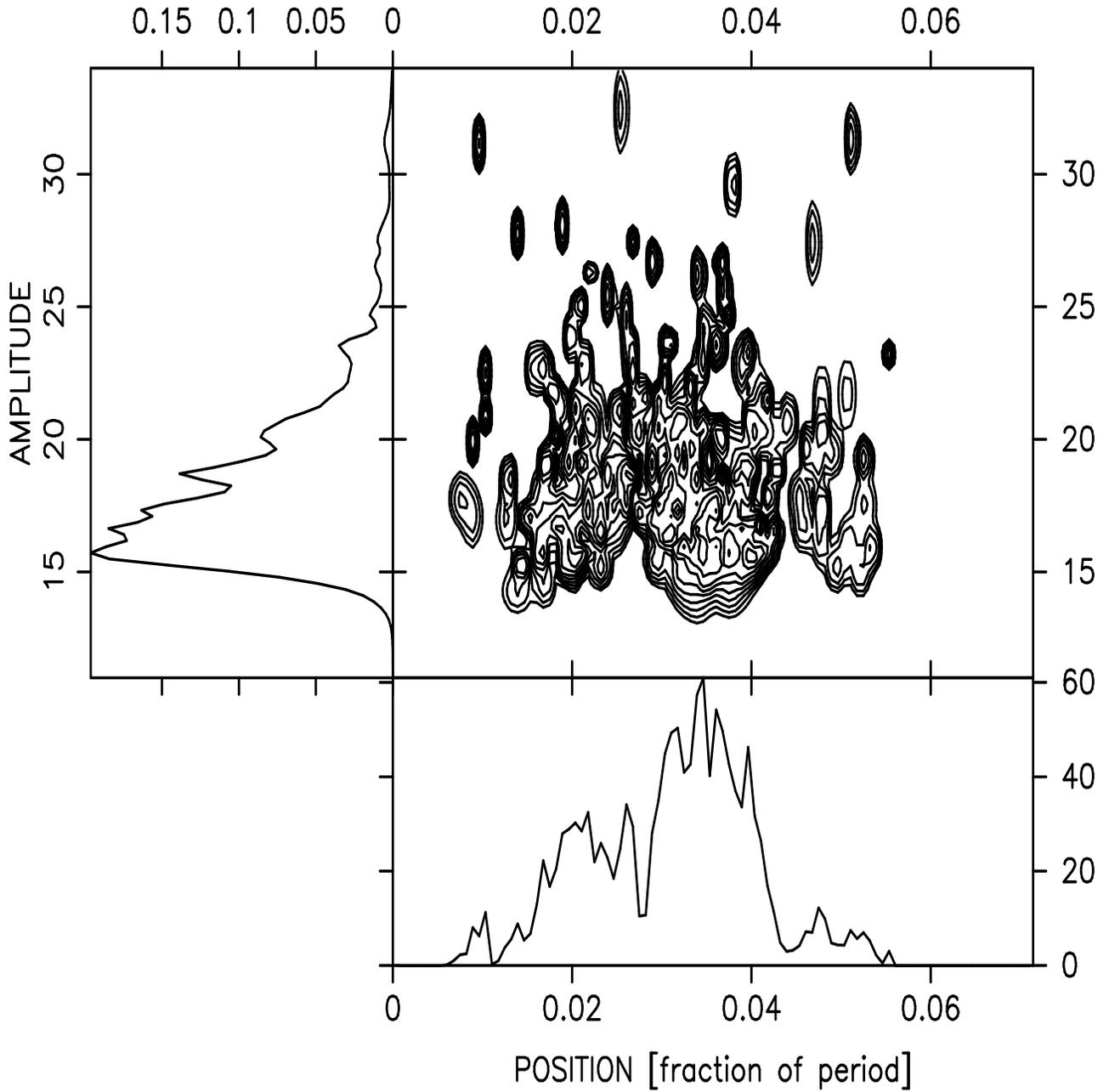}
\end{picture}
\caption{Same as in Figure 5, except that it refers to component 1 of the 
integrated profile shown in Figure 1. For reference the peak of component 1 
lies at 0.036 on the abscissa.}
\end{figure*}

\vfill
\eject
\clearpage
\begin{figure*}
\unitlength=1.0cm
\centering
\begin{picture}(17,17)(0,0)
\psfig{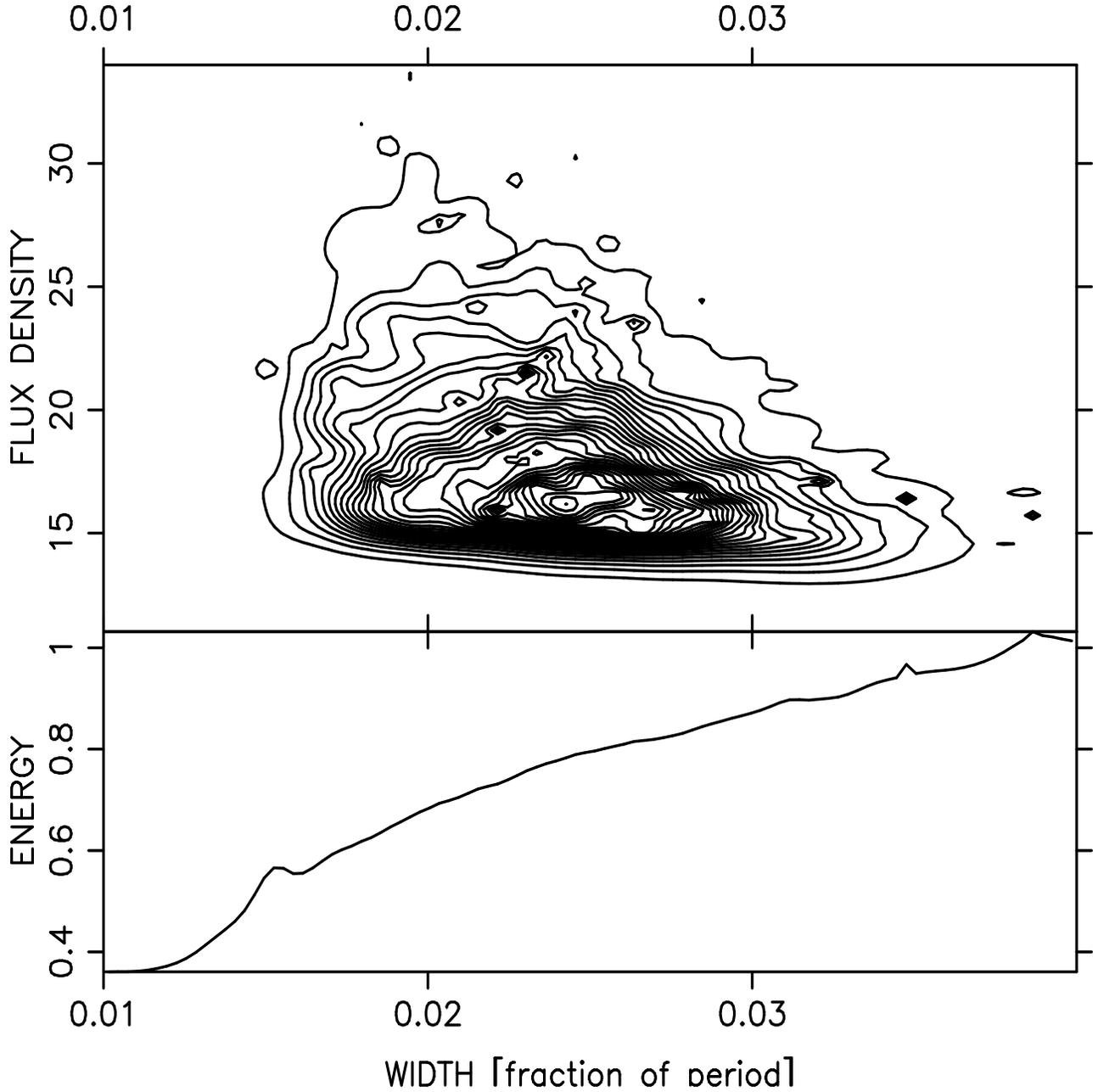}
\end{picture}
\caption{{\bf Top frame}: Probability density function of the peak flux density 
(ordinate, in arbitrary units) of spikes versus their width (abscissa) in component 
2 of the integrated profile shown in Figure 1. A single Gaussian was fit to the 
flux density of each spike. Twenty nine contours are equally spaced in the 
probability range. {\bf Bottom frame}: The average energy in the spike (in arbitrary
units), defined as $\sqrt{\pi}$ times the peak flux density times the width, 
averaged along the ordinate at each value of the abscissa, as a function of the 
width.}
\end{figure*}

\vfill
\eject
\end{document}